\documentclass[aps,preprint,showpacs,floatfix]{revtex4-2}
\usepackage{graphicx,bm,amsmath,dcolumn}
\RequirePackage{color}

\definecolor{MyDarkGreen}{rgb}{0.02,0.60,0.06}

\usepackage[normalem]{ulem}

\begin{document}
	
\title{Exact coefficients of finite-size corrections in the Ising model with Brascamp-Kunz boundary conditions and their relationships for strip and cylindrical geometries}

\author{Nickolay Izmailian}
\email{izmail@yerphi.am}
\affiliation{A. Alikhanyan National Laboratory (Yerevan Physics Institute), Alikhanian Brothers 2, 375036 Yerevan, Armenia}
\author{R. Kenna}
\email{r.kenna@coventry.ac.uk}
\affiliation{Statistical Physics Research Group,
Fluid and
Complex Systems Research Centre,  Coventry University, Coventry, CV1 5FB, England, UK }
\author{Vl.V. Papoyan}
\email{vpap@theor.jinr.ru} \affiliation{Bogoliubov Laboratory of Theoretical Physics, Joint Institute for Nuclear Research, 141980 Dubna, Russia}
\affiliation{Dubna State University, 141980 Dubna, Russia}
\date{\today}

\begin{abstract}
We derive exact finite-size corrections for the free energy $F$ of the Ising model on the ${\cal M} \times 2 {\cal N}$ square lattice with Brascamp-Kunz boundary conditions.
We calculate ratios $r_p(\rho)$ of $p$th coefficients of F for the infinitely long cylinder (${\cal M} \to \infty$) and the infinitely long Brascamp-Kunz strip (${\cal N} \to \infty$) at varying values of the aspect ratio  $\rho={(\cal M}+1) / 2{\cal N}$.
Like previous studies have shown for the two-dimensional dimer model, the limiting values $p \to \infty$ of $r_p(\rho)$  exhibit abrupt anomalous behaviour at certain values of $\rho$.
These critical values of $\rho$ and the limiting values of the  finite-size-expansion-coefficient ratios differ, however, between the two models.

\end{abstract}


\maketitle

\section{Introduction}
\label{introduction}

Introduced by Fisher and Barber in 1972,  finite-size scaling is an essential tool for the study of
thermodynamic properties of systems of any dimensionality \cite{FisherBarber}.
 It enables critical and non-critical properties of infinite systems to be extracted from their finite-size analogues.
 Finite systems necessarily have boundaries and these have to be consistently accounted for even if they play no role in the thermodynamic limit.
 The roles played by boundary conditions are therefore essential in the study of critical phenomena.

Analytical and exact results have key roles in determination of  correct forms for  finite-size scaling. Unlike simulational approaches they are devoid of stochastic error and as such are essential for deeper insights into the roles of boundary conditions.
However, very few models have yielded to analytical or exact approaches. Amongst those who have, the dimer and Ising models in low dimensions are prime examples.

In Ref.\cite{Izmailian2019}, Izmailian, Papoyan and Ziff  recently  considered the dimer model on the rectangular $2M \times 2N$ lattice with free boundary conditions.
In doing so, they built on work by  Kasteleyn \cite{Kasteleyn1961}, as well as by Temperley  and Fisher \cite{TemFi1961}, who
in the 1960's  found the first few coefficients of the asymptotic expansion of the free energy.
This is still considered as one of the most important  exact results of fundamental statistical mechanics.
Izmailian, Papoyan and Ziff derived exact expressions for these coefficients
in terms of the elliptic theta functions $\theta_2, \theta_3, \theta_4$ and the elliptic integral of second kind $E$, up to 22nd order.
A main outcome of their work was an ``intriguing and unexplained simple relation between the
coefficients in the asymptotic series for a square, rectangle, and strip, in the limit of high order'' \cite{Izmailian2019}.

In particular, they considered the dimer model on a $2M \times 2N$ lattice with aspect ratio $\rho=M/N$.
If  $f_p(\rho)$ are the finite-size correction terms [coefficients of powers of  $(MN)^{-{(p+1)}}$] in the  expansion of the free energy, and if their values are   $f_p^\mathrm{sq} = f_p(1)$ for the square lattice,
the ratio
\begin{equation}
 r_p(\rho) = \frac{f_p(\rho)\rho^{-p-1}}{f_p^\mathrm{sq}}
\label{rhorho}
\end{equation}
abruptly jumps from { $r_p=1/2$ when $\rho >1$ to $r_p=1$ at $\rho=1$ } in the infinite $p$ limit.
(Invariance of finite-size scaling correction terms $t_p(\rho)$ under $\rho \to 1/\rho$ ensures a similar relation for $\rho < 1/2$.)
The limit $M \to \infty$ (with finite $N$) delivers a strip geometry and, as shown in Ref.\cite{Izmailian2019}, with
$f^\mathrm{str}_p=\lim_{\rho \to \infty}{f_p(\rho)\rho^{-p-1}}$, one also therefore finds  that $r_p={f_p^\mathrm{str}}/{f_p^\mathrm{sq}}$ in the limit of large $p$   {also reaches these values}.

The existence of an abrupt jump at a specific value of the aspect ratio was described in Ref.\cite{Izmailian2019} as ``very surprising'' and the authors did  ``not have any physical explanation why those ratios should be finite and equal $1/2$'' {(or 1).}
To gain further insight, the authors suggested consideration of other accessible models. One of these is
the Ising model with Brascamp–Kunz (BK) boundary conditions, and that is the subject of this paper.

 The  rectangular lattice Ising model with BK boundary conditions is an iconic example of a model allowing an exact solution \cite{Brascamp}.
Janke and Kenna derived exact expressions for the finite-size scaling of the Fisher zeros, the critical specific heat, the effective critical points, and the specific-heat peak in Ref.\cite{Janke}. At the same time, Izmailian, Oganesyan and Hu analyzed finite-size corrections for the free energy and the specific heat for the model \cite{Izmailian2002}. The approach that was used in this work had been proposed by Ivashkevich Izmailian and Hu in \cite{Ivasho}. It is a systematic method to compute the finite-size corrections for the partition function of free models on a torus, including the dimer
Ising and Gaussian models on a
rectangular lattice. In particular they derive all terms of the exact asymptotic expansion of the logarithm of the partition function on a torus for a class of free exactly solvable models of statistical mechanics. This approach is based on an intimate relation between the terms of the asymptotic expansion and Kronecker’s double series.

The objective of the present paper is to study similar ratios for the ${\cal M} \times 2{\cal N}$ Ising model with BK  boundary conditions,
because of the puzzles posed by Ref.\cite{Izmailian2019}.
In order to match Ref.\cite{Izmailian2019}, we consider explicit calculations up to the 22nd order.
We are especially interested in the
 ${\cal M} \rightarrow \infty$ limit, which gives an infinitely long cylinder, and the ${\cal N} \rightarrow \infty$ limit which delivers an infinitely-long strip.
Our purpose is to find the higher-order asymptotic expansions
in both cases.
We develop two counterparts for Eq.(\ref{rhorho}) , depending on the  aspect ratio $\rho={(\cal M}+1) / 2{\cal N}$.
Rather surprisingly, we find that these
 approach $1$ for increasing $p$ values
 provided $\rho \ne 1/2$, while they vanish precisely at $\rho=1/2$
This abrupt change is similar to that observed in Ref.\cite{Izmailian2019} but its location differs.

\section{
Ising model under Brascamp-Kunz boundary conditions}
\label{First part}

\begin{figure}[!t]
  	\includegraphics[width=70mm]{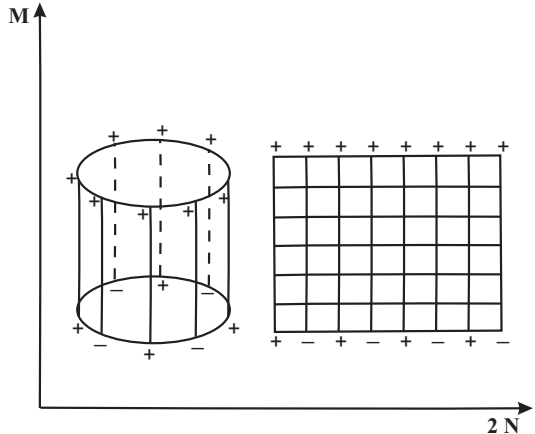}
	\caption{The Brascamp - Kunz boundary conditions.}
	\label{fig0}
\end{figure}

We study the Ising model with
$2 {\cal N}$ sites in the $x$ direction and ${\cal M}$ sites in the $y$ direction and  BK boundary conditions.
These are periodic in the $x$  direction and fixed in the $y$ direction.
In particular, the spins are up (+1) along  the upper border of the resulting cylinder and have  alternating values (+1/-1) along the lower border  as  shown in  Fig.\ \ref{fig0}.
An even number  ($2{\cal N}$) of sites in the $x$ direction ensures an equal number of up and down spins along the lower border.

The Hamiltonian of the two dimensional Ising model is given by
\begin{equation}
\beta H=-J\sum_{<ij>}{s_{i}s_{j}}
\label{hamilton}
\end{equation}
where $s_i$ is the spin variable with two possible values $+1$ and $-1$, the sum runs over the nearest neighbor pairs of spins $<ij>$ and $J$ is exchange energy {which incorporates the inverse temperature }  $\beta=1/k_B T$.

The partition function of the Ising model is given by the sum over
all spin configurations $\{s\}$ on the lattice
\begin{equation}
Z_{^\mathrm{Ising}}(J)=\sum_{{\{s\}}}{e^{-\beta H(s)}}.
\label{Ising}
\end{equation}

For the BK boundary conditions, the Ising partition function on the ${\cal M} \times 2{\cal N}$ rectangular lattice ($Z_{{\cal M},2{\cal N}}$) given in Ref. \cite{Brascamp} can be rewritten as \cite{Izmailian2002}
\begin{equation}
Z_{{\cal M},2{\cal N}}^{2}=\frac{\left(\sqrt{2} e^{\mu}\right)^{4{\cal M}{\cal N}}}
{4\cosh{\left[2 {\cal N} \omega_{\mu}(0)\right]} \cosh{\left[2
		{\cal N} \omega_{\mu}(\pi/2) \right]}} \; Z_{1/2,0}(\mu),
\label{ZtoZmu}
\end{equation}
where $\mu = 1/2 \ln{\sinh{2 J}}$ and
\begin{equation}
\omega_\mu(k)={\rm arcsinh}\sqrt{\sin^2 k+2\,{\rm sinh}^2\mu}
\label{SpectralFunction}
\end{equation}
is a lattice dispersion relation.
Here $Z_{1/2,0}(\mu)$ is the  partition function with twisted boundary conditions
\begin{equation}
Z^2_{1/2,0}(\mu)=
\prod_{i=0}^{N-1}\prod_{j=0}^{M-1}4\left[\textstyle{
	\;\sin^2\left(\frac{\pi (i+1/2)}{N}\right)+
	\sin^2\left(\frac{\pi j}{M}\right)+2\,{\rm sinh}^2\mu
	\;}\right]
\label{Zab2}
\end{equation}
 with $N = 2 {\cal N}$ and $M = 2 ({\cal M}+1)$. The general theory about asymptotic expansion of $Z_{1/2,0}(\mu)$ has been given in Ref. \cite{Ivasho}.

For our further purposes it is useful to  transform
the partition function $Z_{1/2,0}(\mu)$ into the
simpler form
\begin{equation}
Z_{1/2,0}(\mu)=\prod_{n=0}^{N-1} 2 \textstyle{~\!{\rm
sinh}\left[M\omega_\mu\!\left(\frac{\pi(n+1/2)}{N}\right)
\right]}.
\label{Zab}
\end{equation}
\section{Asymptotic Expansion of the Free Energy and ratio of coefficients}
\label{Second part}
The exact asymptotic expansion
of the logarithm of the partition function with twisted boundary
conditions at the critical point ($\mu=\mu_c=0$) can be written as \cite{Ivasho,Izmailian2002}
\begin{eqnarray}
\ln
Z_{1/2,0}(0)&=&\frac{S}{\pi} \int_{0}^{\pi}\!\!\omega_0(x)~\!{\rm d}x  +
\ln\frac{\theta_4}{\eta}-2\pi\xi\sum_{p=1}^{\infty}
\left(\frac{\pi^2\xi}{ S}\right)^{p}\frac{\Lambda_{2p}}{(2p)!}\,
\frac{{\tt Re}\;{\rm K}_{2p+2}^{1/2,0}(i\lambda\xi)}{2p+2}
\label{ExpansionOflnZab}
\end{eqnarray}
Here $S = N M =4 {\cal N}({\cal M}+1)$, $\xi = M/N = ({\cal M}+1)/{\cal N}$,  $\eta=(\theta_2\theta_3\theta_4/2)^{1/3}$ is the
Dedekind-$\eta$ function; $\theta_2$, $\theta_3$, $\theta_4$ are elliptic $\theta$-functions, ${\rm K}_{2p+2}^{1/2,0}(i\lambda\xi)$  are Kronecker's double series \cite{Ivasho, Weil}, $\int_{0}^{\pi}\!\!\omega_0(x)~\!{\rm d}x = 2 G$ and $G = 0.915966\dots$ is  Catalan's constant.

The differential operators $\Lambda_{2p}$ that have appeared here
can be expressed via coefficients $\lambda_{2p}$ of the Taylor expansion of the lattice dispersion relation $\omega_0(k)$
\begin{equation}
\omega_0(k)=k\left(\lambda+\sum_{p=1}^{\infty}
\frac{\lambda_{2p}}{(2p)!}\;k^{2p}\right)
\label{SpectralFunctionExpansion}
\end{equation}
where $\lambda=1$, $\lambda_2=-2/3$, $\lambda_4=4$, etc.
The relations between  differential operators $\Lambda_{2p}$ and  coefficients $\lambda_{2p}$ can be found for example in \cite{Ivasho,Izmailian2019}.

After reaching this point, using Eqs. (\ref{ZtoZmu}), (\ref{SpectralFunction}) and (\ref{ExpansionOflnZab}), one can easily write down the exact asymptotic expansion of the free energy, $F=-\ln{Z_{{\cal M}, 2 {\cal N}}}$, at the critical point.
\begin{equation}
F = {\cal S} f_{\mathrm{bulk}}+2{\cal N}f_{1}+f_0(\rho)
+\sum_{p=1}^{\infty}\frac{f_{p}(\rho)}{{\cal S}^{p}}
\label{freeenergy2}
\end{equation}
Here ${\cal S}$ is the area of the lattice with an additional row and $\rho$ is the aspect ratio
\begin{equation}
{\cal S}=2({\cal M}+1){\cal N} \qquad \mbox{and}  \qquad  \rho = \frac{{\cal M}+1}{2{\cal N}} \label{Srho}
\end{equation}
and
\begin{eqnarray}
f_{bulk}&=&-\frac{1}{2}\ln{2}-\frac{2 G}{\pi}=-0.929695...
\label{fb}\\
f_{1}&=&\frac{1}{2}\ln{2(1+\sqrt{2})}
=0.78726...
\label{f1}\\
f_0(\rho)&=&-\frac{1}{2}\ln{\frac{\theta_4(2\rho)}{2\eta(2\rho)}}=\frac{1}{2}\ln{\frac{\theta_2(\rho)}{\eta(\rho)}}
\label{f0}\\
f_p(\rho)&=&\frac{{\pi}^{2p+1}\rho^{p+1}}{(2p)!(p+1)}\Lambda_{2p}\,
{\tt Re}\;{\rm K}_{2p+2}^{1/2,0}(2i\lambda\rho).
\label{fp}
\end{eqnarray}
Now using expressions for $K_{2p+2}^{\frac{1}{2},0}(\rho)$ which are given in  Supplementary Materials and the expressions for $\Lambda_{2p}$ from \cite{Izmailian2019},  we can express the subleading correction terms $f_p(\rho)$ in the asymptotic expansion of the free energy for the Ising model on the ${\cal M} \times 2{\cal N}$ rectangular lattice with BK boundary conditions for any value of $p$ in terms of the elliptic theta functions ${\theta}_2, {\theta}_3, {\theta}_4$ and the elliptic integral of the second kind $E$. In particular in this paper we have calculated the subleading correction terms $f_p$ in terms of the elliptic functions the elliptic integral of the second kind up to $p=22$. Due to very large expressions for $f_p(\rho)$ for $p > 5$ we have not listed those expressions in Appendix\ \ref{fexpression}.

\begin{figure}[!t]
  	\includegraphics[width=130mm]{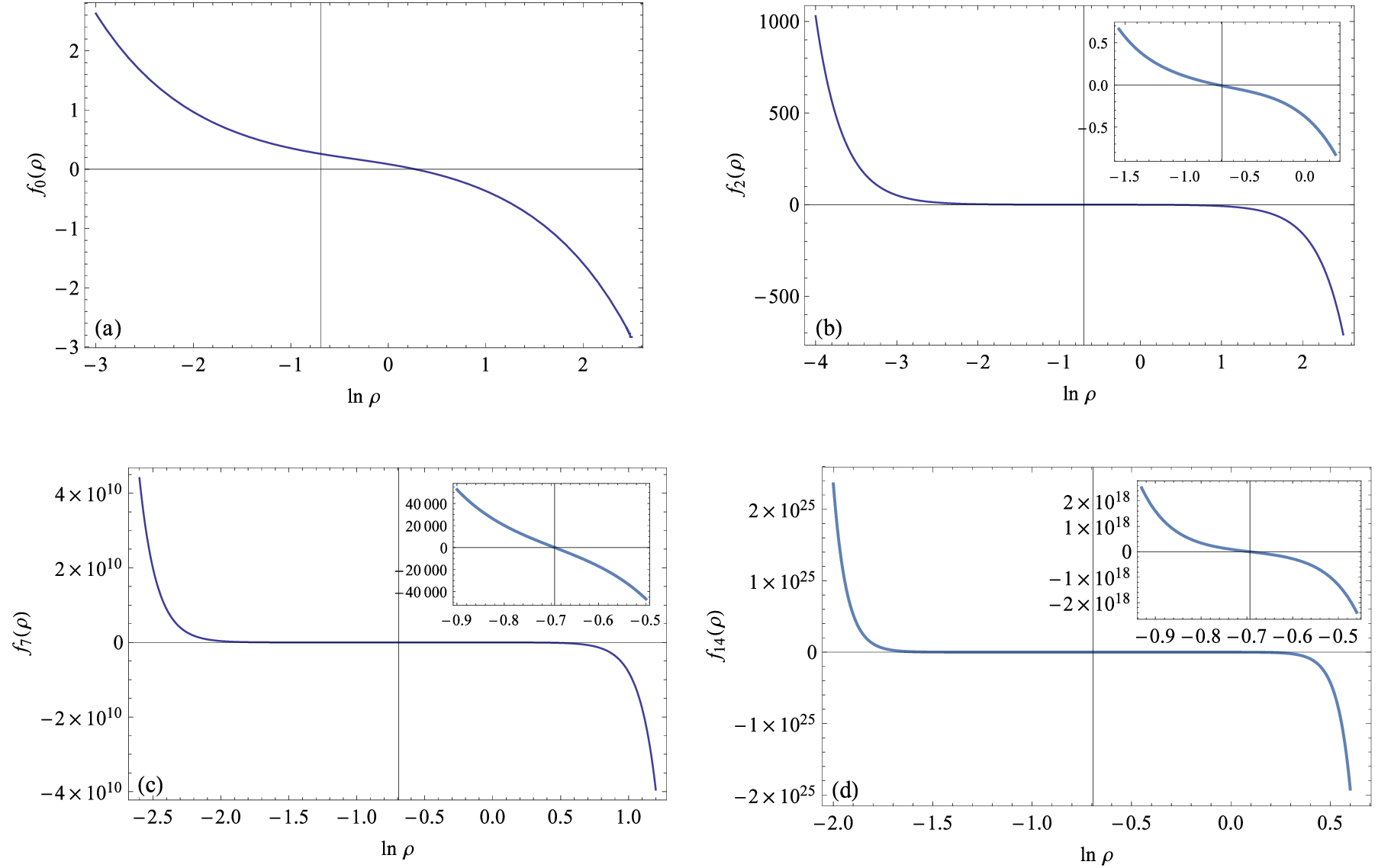}
	\caption{The behavior of the correction
		terms $f_p(\rho)$ for (a) $p=0$, (b) $p=2$, (c) $p=7$, {and} (d) $p=14$.}
	\label{fig1}
\end{figure}
We also obtained the relations between the coefficients for Laurent expansion of the Weierstrass function under transformations $\theta_2 \leftrightarrow \theta_4$ , $\theta_3 \leftrightarrow \theta_4$  and $\theta_2 \leftrightarrow \theta_3$ (see Appendix\ \ref{Weierstrass}).
The relations between Kronecker's double series $K_{2p}^{1/2, 1/2}(\tau)$, $K_{2p}^{1/2, 0}(\tau)$, $K_{2p}^{0, 1/2}(\tau)$ and $K_{2p}^{0, 0}(\tau)$
under the same transformations are proved in Appendix\ \ref{KroneckerToTheta}.

Using the expressions  for subleading correction terms $f_p(\rho)$ in terms of the elliptic theta functions and the elliptic integral and relations with gamma function (see Appendix\ \ref{ThetaToGamma}), we obtain the exact values for $f_p(\rho)$ namely for  $\rho =\frac{1}{4}$, $\rho =\frac{1}{2}$, $\rho =1$  and $\rho =2$ up to $p=22$ (see Supplementary Materials).
Exact numerical values of $f_p(\rho)$ for these particular values of the aspect ratio $\rho$,  up to $p=22$ are given in Table\ \ref{tabular1}.
In Fig.\ \ref{fig1} we plot the behavior of the subleading correction terms $f_p(\rho)$ (a) $p=0$, (b) $p=2$, (c) $p=7$, {and} (d) $p=14$ as a function of the aspect ratio $\rho$.

\begin{figure}[!t]
  	\includegraphics[width=70mm]{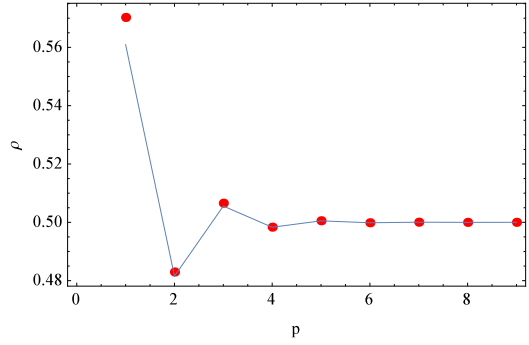}
	\caption{The behavior of roots of the correction
		terms $f_p(\rho)$, as a function of $p$. The dots represent our exact results. The solid line is given by $(-1)^{p+1} a \, b^p + 1/2$, with $a=0.2014$ and $b=0.3027$ and represents the exponential decay to 1/2.}
	\label{fig2}
\end{figure}

We noticed that the roots coefficients $f_p(\rho)$ as a function of the aspect ratio $\rho$ for increasing $p$ exponentially  tends to $1/2$. This is shown in the Fig.\ \ref{fig2}.

Using Kronecker's functions asymptotic form (see for example \cite{Izmailian2007}) when
$\rho \to \infty$  (i.e. ${\cal M} \to \infty$) for fixed $2 {\cal N}$ from Eq. (\ref{freeenergy2}) one obtains the free energy
expansion for infinitely long cylinder of circumference $2 {\cal N}$ with periodic boundary conditions
\begin{eqnarray}
\lim_{{\cal M} \to \infty}\frac{F}{{\cal M}+1}&=&2{\cal N} f_{bulk}-\frac{\pi}{24 {\cal N}}
+\sum_{p=1}^{\infty}\frac{f_p^\mathrm{cyl}}{\left(2{\cal N}\right)^{2p+1}}, \label{cylinder}
\end{eqnarray}
where $f_p^\mathrm{cyl}$ is given by
\begin{eqnarray}
f_p^\mathrm{cyl}=\frac{\pi^{2p+1}\lambda_{2p}B_{2p+2}^{1/2}}{(2p)!(p+1)}.\label{fpcylinder}
\end{eqnarray}
In the limit $ \rho \to 0$ (i.e. ${\cal N} \to \infty$) for fixed ${\cal M}$ we obtain the expansion of free energy of infinitely long strip
with BK boundary condition of the width ${\cal M}+1$
\begin{eqnarray}
\lim_{{\cal N} \to \infty}\frac{F}{2{\cal N}}&=&{\cal M} f_{bulk}+f_1
+\frac{\pi}{24({\cal M}+1)}+\sum_{p=1}^{\infty}\frac{f_p^\mathrm{str}}{\left({\cal M}+1\right)^{2p+1}},
\label{strip}
\end{eqnarray}
where $f_p^\mathrm{str}$ is given by
\begin{eqnarray}
f_p^\mathrm{str}=\frac{\pi^{2p+1}\lambda_{2p}B_{2p+2}^{1/2}}{(2-2^{2p+2})(2p)!(p+1)}.\label{fpstrip}
\end{eqnarray}
We have also used the relation of the Bernoulli numbers $B_n\equiv B_n^{0}$ and $B_n^{1/2}=(2^{1-n}-1)B_n$.
From Eqs. (\ref{freeenergy2}), (\ref{cylinder}) and (\ref{strip}) one can obtain
\begin{eqnarray}
f_p^\mathrm{cyl}&=&\lim_{\rho \to \infty}\rho^{-p-1}f_p(\rho) \label{fpfcylinder}\\
f_p^\mathrm{str}&=&\lim_{\rho \to 0}\rho^{p+1}f_p(\rho). \label{fpfstrip}
\end{eqnarray}
Using known relations for the asymptotic behavior of the $\lambda_{2p}$ \cite{asymptotic}
\begin{equation}
\lambda_{2p}=(-1)^p\frac{2^{2p+5/4}(p/e)^{2p}}{(2p+1)\left[\ln(1+\sqrt{2})\right]^{2p+1/2}}\label{asymptlambda}
\end{equation}
and for the asymptotic behavior of the $B^{1/2}_{2p+2}$,
\begin{equation}
B^{1/2}_{2p+2}=(-1)^{p+1}\frac{2(2p+2)!}{(2\pi)^{2p+2}},\label{asymptB}
\end{equation}
we can derive from Eqs. (\ref{fpcylinder}) and (\ref{fpstrip}) the following formulas for the asymptotic behavior of $f^\mathrm{cyl}_p$
\begin{equation}
f^\mathrm{cyl}_p=-\frac{2^{5/4}(p/e)^{2p}}{\pi\left[\ln(1+\sqrt{2})\right]^{2p+1/2}}\label{asympcylinder}
\end{equation}
and $f^\mathrm{str}_p$
\begin{equation}
f^\mathrm{str}_p=-\frac{2^{1/4}(p/e)^{2p}}{\pi(1-2^{1+2p})\left[\ln(1+\sqrt{2})\right]^{2p+1/2}}.
\label{asympstrip}
\end{equation}
The coefficients of asymptotic expansion of the free energy for the Ising model with BK boundary conditions on an infinitely long strip and for infinitely long cylinder are listed in Supplementary Materials. Exact numerical values of $f^\mathrm{cyl}_p$ and $f^\mathrm{str}_p$ are given in Table\ \ref{tabular2}.


Finally we consider ratios $r_p$ for the Ising model analogous to Eq.(\ref{rhorho}) for the dimer model.
We write the ratio of the coefficients in the free energy expansion $f_p(\rho)$ for $\rho < 1/2$ multiplied by $\rho^{p+1}$ and coefficients $f_p$  for strip $f_p^\mathrm{str}$ as
$$
r_p^\mathrm{str}(\rho)=\frac{\rho^{p+1} f_{p}(\rho)}{f_{p}^\mathrm{str}}
$$
and the ratio of the coefficients in the free energy expansion $f_p(\rho)$ for $\rho > 1/2$ multiplied by $\rho^{-p-1}$ and coefficients $f_p$  infinitely long cylinder $f_p^\mathrm{cyl}$
as
$$
r_p^\mathrm{cyl}(\rho)=\frac{\rho^{-p-1} f_{p}(\rho)}{f_{p}^\mathrm{cyl}}.
$$

\begin{figure}[!t]
  	\includegraphics[width=130mm]{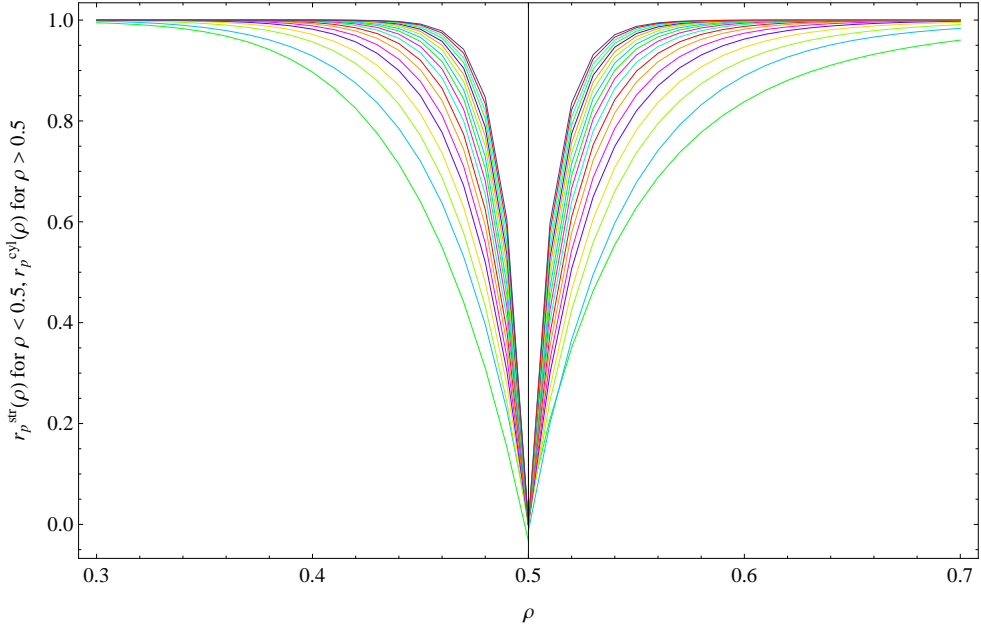}
	\caption{The behavior of ratios of the coefficients $f_p(\rho)$ in the asymptotic expansion of the free energy  with aspect ratio $\rho < 1/2$ times $\rho^{p+1}$  to the asymptotic coefficients for the strip $f_p^\mathrm{str}$ and the coefficients $f_p(\rho)$ with aspect ratio $\rho > 1/2$ times $\rho^{-p-1}$ to the asymptotic coefficients for the cylinder $f_p^\mathrm{cyl}$, as a function of $\rho$ for p =1, 2, 3,...,22 .}
	\label{fig3}
\end{figure}

We list the values of $r_p^\mathrm{str}(\rho)$ and $r_p^\mathrm{cyl}(\rho)$ for $p-0$ up to $p=22$ in Table\ \ref{tabular3} and plot them against $p$ in  Fig.\ \ref{fig3}.
One can clearly see that the ratios $r_p^\mathrm{str}(\rho)$ and $r_p^\mathrm{cyl}(\rho)$  tend to 1 in both cases as $p$ increase and $\rho\neq 1/2$. From  Table\ \ref{tabular4} and  Fig.\ \ref{fig3}, one also sees that the ratio  $r_p$ tends to $0$ for $\rho = 1/2$ and  large values of $p$.

\section{Conclusions}
\label{sec:4}

We have analyzed the partition function of the Ising model with Brascamp-Kunz  boundary conditions on a rectangular $M \times 2N$ lattice.
Using the results of Ref.\ \cite{Izmailian2002} we have calculated the  exact expressions for the correction terms $f_p(\rho)$ in the asymptotic expansion of the free energy from $p=0$ up to $p=22$.
We also found asymptotic formulas of the correction terms as a simple exact expressions for the strip ($N\rightarrow\infty$) and for the infinitely long cylinder ($M\rightarrow\infty$).
We find that the ratio of
the coefficients $f_p(\rho)$ in the free energy expansion of the Ising model with Brascamp-Kunz boundary conditions multiplied by $\rho^{p+1}$ for $\rho>1/2$, and the coefficients $f_p^\mathrm{str}$ in the free energy expansion for the band tends to $1$ with an increase in $p$.
 {{Likewise}} the ratio of the coefficients $f_p(\rho)$ multiplied by $\rho^{-p-1}$ for $\rho <1/2$ and the coefficients $f_p^\mathrm{cyl}$ in the free energy expansion for an infinitely long cylinder
tends to $1$ in the infinite $p$ limit.
These ratios tend to $0$ in both cases for $\rho =1/2$.
Thus we have we have similarities with the dimer case with free boundary conditions in that there is an abrupt change, but the critical values of the aspect ratio at which this occurs differ.
To gain  {further} insight  we plan to study the ratio $r_p(\rho)$ in other models, such as the spanning tree model under different boundary conditions and the dimer model on cylinder.

 \section*{Acknowledgement}
NI acknowledges the receipt of the Armenian SCS grant 20TTWS-1C035.
{NI and RK acknowldge}  support  of  CS  MESRA in the frame of the research project No.SCS 21AG-1C006.


 \section*{Personal statement on behalf of RK}
The work reported in this paper was brought to a conclusion before Putin’s war against Ukraine. In line with protocol of the Middle European Cooperation on Statistical Physics (MECO), of which Kenna is a Board
 Member, Kenna condemns the actions of Russia’s leadership and calls for the withdrawal of Russian forces from Ukraine with immediate effect. Also in line with MECO protocol, and that of Kenna's research group, Kenna suspends future work with Russian  {institutions} until these aims are achieved.
 The full MECO statement about the war is available at
https://sites.google.com/site/mecoconferencephysics/home .

\newpage
\begin{table}[ht]
	\caption{Coefficients $f_p(\rho)$ in the asymptotic expansion of the free energy  for the aspect ratios $\rho = 1/4, 1/2, 1$ and $2$
		.}
	\label{tabular1}
	\begin{center}
		\begin{tabular}{| c | c | c | c | c |}
			\hline
			$p$ & $f_{p}(\frac{1}{4})$ & $f_{p}(\frac{1}{2})$ & $f_{p}(1)$ & $f_{p}(2)$ \\\hline
			0 & 0.523595288249... & 0.259930192710... & 0.086643397569... & -0.177021697969... \\\hline
			1 & 0.172112916479... & 0.0235092640722... & -0.141055584433 & -0.602827737753... \\\hline
			2 & 0.402518667192... & -0.0105273428041... & -0.378984340948 & -3.13653937537 \\\hline
			3 & 3.139139013072... & 0.0204946052696... & -3.12609670831 & $ -5.01013911657...\times 10^{1} $ \\\hline
			4 & $ 4.90234556756...\times 10^{1} $ & -0.0507766330745... & $ -4.89581568797...\times 10^{1} $ & $ -1.56775657752...\times 10^{3}  $ \\\hline
			5 & $ 1.27075751699...\times 10^{3} $ & 0.254955330196... & $ -1.26996097967...\times 10^{3} $ & $ -8.13027369379...\times 10^{4} $ \\\hline
			6 & $ 4.92884549362...\times 10^{4} $ & $ -1.59388671467... $ & $ -4.92817684759...\times 10^{4} $ & $  -6.30849589226...\times 10^6 $ \\\hline
			7 & $ 2.67339869959...\times 10^6 $ & $ 1.54483233041...\times 10^{1} $ & $ -2.67332996787...\times 10^6 $ & $ -6.84381253986...\times 10^8 $ \\\hline
			8 & $ 1.93193808279...\times 10^8 $ & $ -1.81074591531...\times 10^{2} $ & $ -1.93192376909...\times 10^8 $ & $ -9.89148635893...\times 10^{10} $ \\\hline
			9 & $ 1.79407097088...\times 10^{10} $ & $ 2.88511426577...\times 10^{3} $ & $ -1.79406726772...\times 10^{10} $  & $ -1.83712677787...\times 10^{13} $ \\\hline
			10 & $ 2.08177969984...\times 10^{12} $ & $ -5.44134211547...\times 10^{4} $ & $ -2.08177870632...\times 10^{12} $ & $ -4.26348380796...\times 10^{15} $ \\\hline
			11 & $ 2.95162703856...\times 10^{14} $ & $ 1.29349121330...\times 10^6 $ & $ -2.95162670097...\times 10^{14} $ & $ -1.20898636585\times 10^{18} $ \\\hline
			12 & $ 5.02087704137...\times 10^{16} $ & $ -3.57491454591...\times 10^7 $ & $ -5.02087689081...\times 10^{16} $ & $ -4.11310241062...\times 10^{20} $ \\\hline
			13 & $ 1.00920337011...\times 10^{19} $ & $ 1.18780154464...\times 10^9 $ & $ -1.00920336244...\times 10^{19} $ & $ -1.65347879531...\times 10^{23} $ \\\hline
			14 & $ 2.36629829757...\times 10^{21} $ & $ -4.51818179444...\times 10^{10} $ & $ -2.36629829319\times 10^{21} $ & $ -7.75388625431...\times 10^{25} $ \\\hline
			15 & $ 6.40123654627...\times 10^{23} $ & $ 2.00185252421...\times 10^{12} $ & $-6.40123654331...\times 10^{23}$  & $ -4.19511438199...\times 10^{28} $ \\\hline
			16 & $ 1.97885524754...\times 10^{26} $ & $ -1.00199376933...\times 10^{14} $ & $ -1.97885524731...\times 10^{26} $ & $ -2.59372514990...\times 10^{31} $ \\\hline
			17 & $ 6.93253263203...\times 10^{28} $ & $ 5.71313815236...\times 10^{15} $ & $ -6.93253263183...\times 10^{28} $ & $ -1.81732183426...\times 10^{34} $ \\\hline			
			18 & $ 2.73210458826...\times 10^{31} $ & $ -3.63812558184...\times 10^{17} $ & $ -2.73210458824...\times 10^{31} $ & $ -1.43240965036...\times 10^{37}  $ \\\hline
			19 & $ 1.20333420842...\times 10^{34} $ & $ 2.59618103930...\times 10^{19} $ & $ -1.20333420842...\times 10^{34} $ & $ -1.26178737093...\times 10^{40} $ \\\hline
			20 & $ 5.88863561221...\times 10^{36} $ & $ -2.04884884101...\times 10^{21} $ & $ -5.88863561221...\times 10^{36} $ & $ -1.23493639514...\times 10^{43} $ \\\hline
			21 & $ 3.18488175832...\times 10^{39} $ & $ 1.78934250420...\times 10^{23} $ & $ -3.18488175832...\times 10^{39} $ & $ -1.33583622984...\times 10^{46} $ \\\hline
			22 & $1.89474701883...\times 10^{42} $ & $ -1.71361252007...\times 10^{25} $ & $ -1.89474701883\times 10^{42} $ & $ -1.58942900001...\times 10^{49} $ \\\hline
		\end{tabular}
	\end{center}
\end{table}

\begin{table}[ht]
	\caption{Coefficients $f_p$ in the asymptotic expansion of the free energy for the infinite strip ($f_{p}^\mathrm{str}$) and for the infinite cylinder ($f_{p}^\mathrm{cyl}$).}
	\label{tabular2}
	\begin{center}
		\begin{tabular}{| c | c | c |}
			\hline
			$p$ & $f_{p}^\mathrm{str}$ & $f_{p}^\mathrm{cyl}$  \\\hline
			0 & 0.130899693900... & -0.261799387799... \\\hline
			1 & 0.0107660682918... & -0.150724956085... \\\hline
			2 & 0.00632481160684... & -0.392138319624... \\\hline
			3 & 0.0123286192230... & -3.13146928264349... \\\hline
			4 & 0.0479378812728... & $ -4.89925146608...\times 10^{1} $ \\\hline
			5 & 0.310296854792... & $ -1.27035532352...\times 10^{3} $ \\\hline
			6 & 3.00849250297... & $ -4.92851241836...\times 10^{4} $ \\\hline
			7 & $ 4.07935464563...\times 10^{1} $ & $ -2.67336427347...\times 10^6 $ \\\hline
			8 & $ 7.36978786113...\times 10^{2} $ & $ -1.93193092949...\times 10^8 $ \\\hline
			9 & $ 1.71096090406...\times 10^{3} $ & $ -1.79406911902...\times 10^{10} $ \\\hline
			10 & $ 4.96335076279...\times 10^{5} $ & $ -2.08177920311...\times 10^{12} $ \\\hline
			11 & $ 1.75930692054...\times 10^7 $ & $ -2.95162686976...\times 10^{14} $ \\\hline
			12 & $ 7.48168992359...\times 10^8 $ & $ -5.02087696609...\times 10^{16} $ \\\hline
			13 & $ 3.75957553757...\times 10^{10} $ & $ -1.00920336628...\times 10^{19} $ \\\hline
			14 & $ 2.20378702487...\times 10^{12} $ & $ -2.36629829538...\times 10^{21} $ \\\hline
			15 & $ 1.49040402560...\times 10^{14} $ & $ -6.40123654479...\times 10^{23} $ \\\hline
			16 & $ 1.15184535252...\times 10^{16} $ & $ -1.97885524742...\times 10^{26} $ \\\hline
			17 & $ 1.00881627181...\times 10^{18} $ & $ -6.93253263193...\times 10^{28} $ \\\hline			
			18 & $ 9.93933859088...\times 10^{19} $ & $ -2.73210458825...\times 10^{31} $ \\\hline
			19 & $ 1.09442608702...\times 10^{22} $ & $ -1.20333420842...\times 10^{34} $ \\\hline
			20 & $ 1.33892072249...\times 10^{24} $ & $ -5.88863561221...\times 10^{36} $ \\\hline
			21 & $ 1.81039567810...\times 10^{26} $ & $ -3.18488175832...\times 10^{39} $ \\\hline
			22 & $ 2.69259746066...\times 10^{28} $ & $ -1.89474701883...\times 10^{42} $ \\\hline
		\end{tabular}
	\end{center}
\end{table}

\begin{table}[ht]
	\caption{Ratios of the coefficients in the asymptotic expansion of the free energy $f_p(\rho)$ with $\rho = 1/4$ times $(\frac{1}{4})^{p+1}$ to the asymptotic coefficients for the strip $f_p^\mathrm{str}$, $f_p(\rho)$ with $\rho = 1$ to the asymptotic coefficients for the cylinder $f_p^\mathrm{cyl}$ and $f_p(\rho)$ with $\rho = 2$ times $2^{-p-1}$ to the asymptotic coefficients for the cylinder $f_p^\mathrm{cyl}$ as a function of $p$.}
	\label{tabular3}
	\begin{center}
		\begin{tabular}{| c | c | c | c |}
			\hline
			$p$ &  $r_p^\mathrm{str}=\frac{(\frac{1}{4})^{p+1} f_{p}(\frac{1}{4})}{f_{p}^\mathrm{str}}$ &  $r_p^\mathrm{cyl}=\frac{f_p(1)}{f_{p}^\mathrm{cyl}}$& $r_p^\mathrm{cyl}=\frac{2^{-p-1} f_p(2)}{f_{p}^\mathrm{cyl}}$ \\\hline
			0 & 0.999993339655095528017488 & -0.3309534002289773900946846 & 0.3380865391971407478281185 \\\hline
			1 & 0.999163017402809561962496 & 0.935847573601313135868700 & 0.999880433559350175061247 \\\hline
			2 & 0.994393914923976306061227 & 0.966455768238489643610306 & 0.999819202307537384224738 \\\hline
			3 & 0.994617608671787826597893 & 0.998284327945495711605871 & 0.999957740353741161966976 \\\hline
			4 & 0.998677187268516332817291 & 0.999298713663337705289882 & 0.999997517714294504507257 \\\hline
			5 & 0.999828163357488075062333 & 0.999689579882906199027829 & 0.999999953663715280854808 \\\hline
			6 & 0.999945502733509412893116 & 0.999931912362528598890457 & 0.999999999486273115273625 \\\hline
			7 & 0.999982359481393231263486 & 0.999987167628971390721966 & 0.999999999968052918721302 \\\hline
			8 & 0.999996073244944444713617 & 0.999996293656869882221401 & 0.999999999992907881683022 \\\hline
			9 & 0.999999124863209607040903 & 0.999998968098210751738736 & 0.999999999998966062139553 \\\hline
			10 & 0.999999761773952596695354 & 0.999999761363057705831311 & 0.999999999999903808483912 \\\hline
			11 & 0.999999937978850492536281 & 0.999999942813981074113102 & 0.999999999999994070123664 \\\hline
			12 & 0.999999985190171839415830 & 0.999999985007330984754956 & 0.999999999999999744620307 \\\hline
			13 & 0.999999996345359234156146 & 0.999999996204074490088418 & 0.999999999999999991073663 \\\hline
			14 & 0.999999999063016580968864 & 0.999999999074337101836494 & 0.999999999999999999579251 \\\hline
			15 & 0.999999999765225046712899 & 0.999999999769113761235235 & 0.999999999999999999963030 \\\hline
			16 & 0.999999999942044723146757 & 0.999999999941539947286117 & 0.999999999999999999996453 \\\hline
			17 & 0.999999999985497727976717 & 0.999999999985398442194306 & 0.999999999999999999999719 \\\hline
			18 & 0.999999999996352260731897 & 0.999999999996371781603060 & 0.999999999999999999999982 \\\hline
			19 & 0.999999999999089370892081 & 0.999999999999091639708485 & 0.999999999999999999999999 \\\hline
			20 & 0.999999999999772972174462 & 0.999999999999772280474320 & 1.00000000000000000000000 \\\hline
			21 & 0.999999999999943177574113 & 0.999999999999943135588192 & 1.00000000000000000000000 \\\hline
			22 & 0.999999999999985777662470 & 0.999999999999985800628097 & 1.00000000000000000000000 \\\hline
		\end{tabular}
	\end{center}
\end{table}

\begin{table}[ht]
	\caption{Ratios of the coefficients $f_p(\rho)$ in the asymptotic expansion of the free energy with aspect ratio $\rho=1/2$ times $(\frac{1}{2})^{p+1}$ to the asymptotic coefficients for the strip $f_p^\mathrm{str}$  and times $(\frac{1}{2})^{-p-1}$ to the asymptotic coefficients for the cylinder $f_p^\mathrm{cyl}$ , as a function of $p$.}
	\label{tabular4}
	\begin{center}
		\begin{tabular}{| c | c | c |}
			\hline
			$p$ &  $r_p=\frac{(\frac{1}{2})^{p+1} f_{p}(\frac{1}{2})}{f_{p}^\mathrm{str}}$ & $r_p=\frac{(\frac{1}{2})^{-p-1} f_{p}(\frac{1}{2})}{f_{p}^\mathrm{cyl}}$\\\hline
			0  &  0.992860200686932170284054 & -1.985720401373864340568108 \\\hline
			1  &  0.545911084600765995923408 & -0.623898382400875423912467 \\\hline
			2  & -0.208056450106897076054997 &  0.214767948497442143024513 \\\hline
			3  &  0.103897509216542038863929 & -0.104715599840294338382543 \\\hline
			4  & -0.033100540563053800834474 &  0.033165316571983456022017 \\\hline
			5  &  0.012838277194207318041263 & -0.012844548946622661137522 \\\hline
			6  & -0.004139029745310480389337 &  0.004139535059648816426498 \\\hline
			7  &  0.001479278418986346793646 & -0.001479323564358794266616 \\\hline
			8  & -0.000479879907330006230188 &  0.000479883568551079770531 \\\hline
			9  &  0.000164673219211188880834 & -0.000164673533301027482952 \\\hline
			10 & -0.000053530476824995503296 &  0.000053530502350328121212 \\\hline
			11 &  0.000017949895470759916822 & -0.000017949897610554458482 \\\hline
			12 & -0.000005832785643838021571 &  0.000005832785817668584927 \\\hline
			13 &  0.000001928346754751898939 & -0.000001928346769119201961 \\\hline
			14 & -0.000000625668289806867818 &  0.000000625668290972265825 \\\hline
			15 &  0.000000204950099949109278 & -0.000000204950100044546606 \\\hline
			16 & -0.000000066368334667218326 &  0.000000066368334674944617 \\\hline
			17 &  0.000000021603430770326987 & -0.000000021603430770955729 \\\hline
			18 & -0.000000006981524767547933 &  0.000000006981524767598730 \\\hline
			19 &  0.000000002262291814200193 & -0.000000002262291814204308 \\\hline
			20 & -0.000000000729667740979856 &  0.000000000729667740980188 \\\hline
			21 &  0.000000000235645998572673 & -0.000000000235645998572700 \\\hline
			22 & -0.000000000075866717571843 &  0.000000000075866717571846 \\\hline	
		\end{tabular}
	\end{center}
\end{table}
\newpage

\appendix
\section{Expressions for $f_p(\rho)$ from $p=0$ up to $p=5$.}
\label{fexpression}
\begin{eqnarray}
f_0 &=& \frac{1}{12} \left(\ln 4-3 \ln \left(\theta _3 \theta _4\right)+2 \ln
   \left(\sqrt{\theta _3 \theta _4} \sqrt{\theta _3^2-\theta _4^2}
   \sqrt[4]{\theta _2^4+2 \theta _4^2 \left(\theta _3^2+\theta
   _4^2\right)}\right)\right) \cr
f_1 &=& \frac{\pi ^3 \rho ^2 \left(\theta _2^8-14 \theta _4^4 \theta _2^4-14
   \theta _4^8\right)}{2880} \cr
f_2 &=& -\frac{\pi ^5 \rho ^4}{48384} \left(-124 \text{E} \theta _3^2 \theta
   _4^{12}-186 \text{E} \theta _2^4 \theta _3^2 \theta _4^8-66
   \text{E} \theta _2^8 \theta _3^2 \theta _4^4-2 \text{E}
   \theta _2^{12} \theta _3^2+62 \pi  \theta _4^{16}+124 \pi
   \theta _2^4 \theta _4^{12}\right.\cr
   &+& \left.73 \pi  \theta _2^8 \theta _4^8+11
   \pi  \theta _2^{12} \theta _4^4\right)-\frac{\pi ^5 \rho ^3}
   {48384}\left(\theta _2^{12}+33 \theta _4^4 \theta _2^8+93 \theta _4^8
   \theta _2^4+62 \theta _4^{12}\right) \cr
f_3 &=& \frac{\pi ^7 \rho ^6}{58060800} \left(-213360 \text{E}^2 \theta
   _4^{20}-640080 \text{E}^2 \theta _2^4 \theta _4^{16}-663600
   \text{E}^2 \theta _2^8 \theta _4^{12}-260400 \text{E}^2
   \theta _2^{12} \theta _4^8\right.\cr
   &-& \left.22680 \text{E}^2 \theta _2^{16}
   \theta _4^4+70 \theta _2^{20} \left(12 \text{E}^2-5 \pi ^2
   \theta _4^4\right)+213360 \pi  \text{E} \theta _3^2 \theta
   _4^{20}+533400 \pi  \text{E} \theta _2^4 \theta _3^2 \theta
   _4^{16}\right.\cr
   &+& \left.438480 \pi  \text{E} \theta _2^8 \theta _3^2 \theta
   _4^{12}+124320 \pi  \text{E} \theta _2^{12} \theta _3^2 \theta
   _4^8+5880 \pi  \text{E} \theta _2^{16} \theta _3^2 \theta
   _4^4-53340 \pi ^2 \theta _4^{24}\right.\cr
   &-& \left.160020 \pi ^2 \theta _2^4
   \theta _4^{20}-173180 \pi ^2 \theta _2^8 \theta _4^{16}-79660
   \pi ^2 \theta _2^{12} \theta _4^{12}-13510 \pi ^2 \theta
   _2^{16} \theta _4^8\right)\cr
   &+& \frac{\pi ^7 \rho ^5}{58060800}
   \left(213360 \text{E} \theta _3^2 \theta _4^{16}+426720
   \text{E} \theta _2^4 \theta _3^2 \theta _4^{12}+236880
   \text{E} \theta _2^8 \theta _3^2 \theta _4^8+23520 \text{E}
   \theta _2^{12} \theta _3^2 \theta _4^4-840 \text{E} \theta
   _2^{16} \theta _3^2\right.\cr
   &-& \left.106680 \pi  \theta _4^{20}-266700 \pi
   \theta _2^4 \theta _4^{16}-219240 \pi  \theta _2^8 \theta
   _4^{12}-62160 \pi  \theta _2^{12} \theta _4^8-2940 \pi  \theta
   _2^{16} \theta _4^4\right) \cr
   &+& \frac{\pi ^7\rho ^4}{58060800} \left(237
   \theta _2^{16}-6636 \theta _4^4 \theta _2^{12}-66834 \theta
   _4^8 \theta _2^8-120396 \theta _4^{12} \theta _2^4-60198 \theta
   _4^{16}\right)  \cr
f_4 &=& \frac{\pi ^9 \rho ^8}{3065610240}\left(-3934700 \pi ^3 \theta _4^{32}-15738800 \pi ^3
   \theta _2^4 \theta _4^{28}+23608200 \pi ^2 \text{E}  \theta
   _3^2 \theta _4^{28}\right.\cr
   &-& \left.47216400 \pi \text{E}^2  \theta
   _4^{28}-24962245 \pi ^3 \theta _2^8 \theta _4^{24}-188865600
   \pi\text{E}^2   \theta _2^4 \theta _4^{24}+82628700 \pi ^2  \text{E}
   \theta _2^4 \theta _3^2 \theta _4^{24}\right.\cr
   &+& \left.31477600
   \text{E}^3 \theta _3^2 \theta _4^{24}-19800935 \pi ^3 \theta
   _2^{12} \theta _4^{20}-293684160 \pi \text{E}^2   \theta _2^8
   \theta _4^{20}+109942140 \pi ^2 \text{E} \theta _2^8 \theta
   _3^2 \theta _4^{20}\right.\cr
   &+& \left.110171600 \text{E}^3 \theta _2^4 \theta
   _3^2 \theta _4^{20}-8099630 \pi ^3 \theta _2^{16} \theta
   _4^{16}-220022880  \pi\text{E}^2  \theta _2^{12} \theta
   _4^{16}+68283600 \pi ^2\text{E}  \theta _2^{12} \theta _3^2
   \theta _4^{16}\right.\cr
   &+& \left.142696400 \text{E}^3 \theta _2^8 \theta _3^2
   \theta _4^{16}-1559635 \pi ^3 \theta _2^{20} \theta
   _4^{12}-78549240\pi \text{E}^2   \theta _2^{16} \theta
   _4^{12}+19242300 \pi ^2\text{E}  \theta _2^{16} \theta _3^2
   \theta _4^{12}\right.\cr
   &+& \left.81312000 \text{E}^3 \theta _2^{12} \theta _3^2
   \theta _4^{12}-103180 \pi ^3 \theta _2^{24} \theta
   _4^8-10736880\pi \text{E}^2   \theta _2^{20} \theta
   _4^8+1894200 \pi ^2\text{E}  \theta _2^{20} \theta _3^2 \theta
   _4^8\right.\cr
   &+& \left.17894800 \text{E}^3 \theta _2^{16} \theta _3^2 \theta
   _4^8-385 \pi ^3 \theta _2^{28} \theta _4^4-175560 \pi\text{E}^2
     \theta _2^{24} \theta _4^4+13860 \pi ^2 \text{E}  \theta
   _2^{24} \theta _3^2 \theta _4^4\right.\cr
   &+& \left.616000 \text{E}^3 \theta
   _2^{20} \theta _3^2 \theta _4^4+30800 \text{E}^3 \theta
   _2^{24} \theta _3^2\right) \cr
     &+& \frac{\pi ^9 \rho ^7}{3065610240}
   \left(-11804100 \pi ^2 \theta _4^{28}-41314350 \pi ^2 \theta
   _2^4 \theta _4^{24}-47216400 \text{E}^2 \theta
   _4^{24}+47216400  \pi\text{E}  \theta _3^2 \theta
   _4^{24}\right.\cr
   &-& \left.54971070 \pi ^2 \theta _2^8 \theta _4^{20}-165257400
   \text{E}^2 \theta _2^4 \theta _4^{20}+141649200 \pi\text{E}
   \theta _2^4 \theta _3^2 \theta _4^{20}-34141800 \pi ^2 \theta
   _2^{12} \theta _4^{16}\right.\cr
   &-& \left.214044600 \text{E}^2 \theta _2^8 \theta
   _4^{16}+152034960 \pi \text{E}   \theta _2^8 \theta _3^2 \theta
   _4^{16}-9621150 \pi ^2 \theta _2^{16} \theta _4^{12}-121968000
   \text{E}^2 \theta _2^{12} \theta _4^{12}\right.\cr
   &+& \left.67987920 \pi \text{E}
    \theta _2^{12} \theta _3^2 \theta _4^{12}-947100 \pi ^2
   \theta _2^{20} \theta _4^8-26842200 \text{E}^2 \theta _2^{16}
   \theta _4^8+10561320 \pi  \text{E} \theta _2^{16} \theta _3^2
   \theta _4^8\right.\cr
   &-& \left.6930 \pi ^2 \theta _2^{24} \theta _4^4-924000
   \text{E}^2 \theta _2^{20} \theta _4^4+175560 \pi\text{E}
   \theta _2^{20} \theta _3^2 \theta _4^4-46200 \text{E}^2 \theta
   _2^{24}\right)\cr
     &+& \frac{\pi ^9 \rho ^6}{3065610240}
   \left(-12907860 \pi  \theta _4^{24}-38723580 \pi  \theta _2^4
   \theta _4^{20}+25815720 \text{E} \theta _3^2 \theta
   _4^{20}-41562804 \pi  \theta _2^8 \theta _4^{16}\right.\cr
   &+& \left.64539300
   \text{E} \theta _2^4 \theta _3^2 \theta _4^{16}-18586308 \pi
   \theta _2^{12} \theta _4^{12}+52490280 \text{E} \theta _2^8
   \theta _3^2 \theta _4^{12}-2887218 \pi  \theta _2^{16} \theta
   _4^8\right.\cr
   &+& \left.14196120 \text{E} \theta _2^{12} \theta _3^2 \theta
   _4^8-47994 \pi  \theta _2^{20} \theta _4^4+479940 \text{E}
   \theta _2^{16} \theta _3^2 \theta _4^4+25260 \text{E} \theta
   _2^{20} \theta _3^2\right)\cr
     &+& \frac{\pi ^9 \rho ^5}{3065610240}
   \left(-4930 \theta _2^{20}-93670 \theta _4^4 \theta
   _2^{16}-2770660 \theta _4^8 \theta _2^{12}-10244540 \theta
   _4^{12} \theta _2^8\right.\cr
   &-& \left.12596150 \theta _4^{16} \theta _2^4-5038460
   \theta _4^{20}\right)  \cr
f_5 &=& \frac{\pi ^{11} \rho
   ^{10}}{167382319104000} \left(-495562016950 \pi ^4 \theta
   _4^{40}-2477810084750 \pi ^4 \theta _2^4 \theta
   _4^{36}+3964496135600 \pi ^3\text{E}  \theta _3^2 \theta
   _4^{36}\right.\cr
   &-& \left.11893488406800 \pi ^2 \text{E}^2  \theta
   _4^{36}-5172413070825 \pi ^4 \theta _2^8 \theta
   _4^{32}-7928992271200 \text{E}^4 \theta _4^{32}\right.\cr
   &-& \left.59467442034000
  \pi ^2 \text{E}^2  \theta _2^4 \theta _4^{32}+17840232610200
  \pi ^3 \text{E}  \theta _2^4 \theta _3^2 \theta
   _4^{32}+15857984542400\pi \text{E}^3   \theta _3^2 \theta
   _4^{32}\right.\cr
   &-& \left.5822791774800 \pi ^4 \theta _2^{12} \theta
   _4^{28}-122650661133000 \pi ^2\text{E}^2  \theta _2^8 \theta
   _4^{28}-39644961356000 \text{E}^4 \theta _2^4 \theta
   _4^{28}\right.\cr
   &+& \left.32706893419200 \pi ^3\text{E}  \theta _2^8 \theta _3^2
   \theta _4^{28}+71360930440800  \pi \text{E}^3  \theta _2^4
   \theta _3^2 \theta _4^{28}-3805057105850 \pi ^4 \theta _2^{16}
   \theta _4^{24}\right.\cr
   &-& \left.133797992328000 \text{E}^2 \pi ^2 \theta
   _2^{12} \theta _4^{24}-80775174079600 \text{E}^4 \theta _2^8
   \theta _4^{24}+31219708119600 \pi ^3\text{E}  \theta _2^{12}
   \theta _3^2 \theta _4^{24}\right.\cr
   &+& \left.128844211496000 \text{E}^3 \pi
   \theta _2^8 \theta _3^2 \theta _4^{24}-1442202461700 \pi ^4
   \theta _2^{20} \theta _4^{20}-81991258549200 \pi ^2 \text{E}^2
   \theta _2^{16} \theta _4^{20}\right.\cr
   &-& \left.85230928182400 \text{E}^4 \theta
   _2^{12} \theta _4^{20}+16396161381600 \text{E} \pi ^3 \theta
   _2^{16} \theta _3^2 \theta _4^{20}+117937064845600 \pi  \text{E}^3
    \theta _2^{12} \theta _3^2 \theta _4^{20}\right.\cr
   &-& \left.296985388700 \pi
   ^4 \theta _2^{24} \theta _4^{16}-27444059042400 \pi
   ^2 \text{E}^2 \theta _2^{20} \theta _4^{16}-48562136823200 \text{E}^4
   \theta _2^{16} \theta _4^{16}\right.\cr
   &+& \left.4568342578800 \pi ^3 \text{E}
   \theta _2^{20} \theta _3^2 \theta _4^{16}+56489240007200
   \pi \text{E}^3   \theta _2^{16} \theta _3^2 \theta
   _4^{16}-27936909000 \pi ^4 \theta _2^{28} \theta
   _4^{12}\right.\cr
   &-& \left.4403680881600  \pi ^2\text{E}^2 \theta _2^{24} \theta
   _4^{12}-13886999526400 \text{E}^4 \theta _2^{20} \theta
   _4^{12}+582531149200 \pi ^3\text{E}  \theta _2^{24} \theta
   _3^2 \theta _4^{12}\right.\cr
   &+& \left.12889275599200 \pi \text{E}^3   \theta
   _2^{20} \theta _3^2 \theta _4^{12}-723998275 \pi ^4 \theta
   _2^{32} \theta _4^8-230037007200\pi ^2 \text{E}^2  \theta
   _2^{28} \theta _4^8\right.\cr
   &-& \left.1500316417600 \text{E}^4 \theta _2^{24}
   \theta _4^8+21819798000\pi ^3 \text{E}  \theta _2^{28} \theta
   _3^2 \theta _4^8+997723927200 \pi\text{E}^3   \theta _2^{24}
   \theta _3^2 \theta _4^8\right.\cr
   &-& \left.350350 \pi ^4 \theta _2^{36} \theta
   _4^4-441441000 \pi ^2\text{E}^2  \theta _2^{32} \theta
   _4^4-1793792000 \text{E}^4 \theta _2^{28} \theta _4^4+21021000
   \pi ^3\text{E}  \theta _2^{32} \theta _3^2 \theta
   _4^4\right.\cr
   &+& \left.1889087200\pi \text{E}^3   \theta _2^{28} \theta _3^2
   \theta _4^4+1936734800 \text{E}^4 \theta _2^{32}\right)\cr
   &+&
   \frac{\pi ^{11} \rho ^9}{167382319104000}\cr
   & & \left(-1982248067800
   \pi ^3 \theta _4^{36}-8920116305100 \pi ^3 \theta _2^4 \theta
   _4^{32}+11893488406800 \pi ^2 \text{E} \theta _3^2 \theta
   _4^{32}\right.\cr
   &-& \left.23786976813600\pi \text{E}^2   \theta
   _4^{32}-16353446709600 \pi ^3 \theta _2^8 \theta
   _4^{28}-107041395661200 \pi \text{E}^2  \theta _2^4 \theta
   _4^{28}\right.\cr
   &+& \left.47573953627200\pi ^2 \text{E}  \theta _2^4 \theta _3^2
   \theta _4^{28}+15857984542400 \text{E}^3 \theta _3^2 \theta
   _4^{28}-15609854059800 \pi ^3 \theta _2^{12} \theta
   _4^{24}\right.\cr
   &-& \left.193266317244000 \pi \text{E}^2   \theta _2^8 \theta
   _4^{24}+75076707505800 \pi ^2 \text{E}  \theta _2^8 \theta _3^2
   \theta _4^{24}+63431938169600 \text{E}^3 \theta _2^4 \theta
   _3^2 \theta _4^{24}\right.\cr
   &-& \left.8198080690800 \pi ^3 \theta _2^{16} \theta
   _4^{20}-176905597268400\pi \text{E}^2   \theta _2^{12} \theta
   _4^{20}+58721284822200 \pi ^2\text{E}  \theta _2^{12} \theta
   _3^2 \theta _4^{20}\right.\cr
   &+& \left.98118409989600 \text{E}^3 \theta _2^8
   \theta _3^2 \theta _4^{20}-2284171289400 \pi ^3 \theta _2^{20}
   \theta _4^{16}-84733860010800\pi \text{E}^2   \theta _2^{16}
   \theta _4^{16}\right.\cr
   &+& \left.23269973727000 \text{E} \pi ^2 \theta _2^{16}
   \theta _3^2 \theta _4^{16}+72343446375200 \text{E}^3 \theta
   _2^{12} \theta _3^2 \theta _4^{16}-291265574600 \pi ^3 \theta
   _2^{24} \theta _4^{12}\right.\cr
   &-& \left.19333913398800 \pi \text{E}^2   \theta
   _2^{20} \theta _4^{12}+4174085315400  \pi ^2\text{E} \theta
   _2^{20} \theta _3^2 \theta _4^{12}+24780827271200 \text{E}^3
   \theta _2^{16} \theta _3^2 \theta _4^{12}\right.\cr
   &-& \left.10909899000 \pi ^3
   \theta _2^{28} \theta _4^8-1496585890800\pi \text{E}^2
   \theta _2^{24} \theta _4^8+229595566200 \pi ^2 \text{E}  \theta
   _2^{24} \theta _3^2 \theta _4^8\right.\cr
   &+& \left.2993171781600 \text{E}^3
   \theta _2^{20} \theta _3^2 \theta _4^8-10510500 \pi ^3 \theta
   _2^{32} \theta _4^4-2833630800 \pi \text{E}^2  \theta _2^{28}
   \theta _4^4+441441000\pi ^2  \text{E} \theta _2^{28} \theta
   _3^2 \theta _4^4\right.\cr
   &+& \left.7461053600 \text{E}^3 \theta _2^{24} \theta
   _3^2 \theta _4^4-3873469600 \text{E}^3 \theta _2^{28} \theta
   _3^2\right)\cr
   &+&
   \frac{\pi ^{11}\rho
   ^8}{167382319104000}
   \left(-3191823929580 \pi ^2 \theta _4^{32}-12767295718320 \pi
   ^2 \theta _2^4 \theta _4^{28}-12767295718320 \text{E}^2 \theta
   _4^{28}\right.\cr
   &+& \left.12767295718320 \pi\text{E}   \theta _3^2 \theta
   _4^{28}-20148136810230 \pi ^2 \theta _2^8 \theta
   _4^{24}-51069182873280 \text{E}^2 \theta _2^4 \theta
   _4^{24}\right.\cr
   &+& \left.44685535014120  \pi \text{E} \theta _2^4 \theta _3^2
   \theta _4^{24}-15758875416570 \pi ^2 \theta _2^{12} \theta
   _4^{20}-78995332124280 \text{E}^2 \theta _2^8 \theta
   _4^{20}\right.\cr
   &+& \left.59047202608680\pi \text{E}   \theta _2^8 \theta _3^2
   \theta _4^{20}-6244901112450 \pi ^2 \theta _2^{16} \theta
   _4^{16}-58243856316360 \text{E}^2 \theta _2^{12} \theta
   _4^{16}\right.\cr
   &+& \left.35904168986400 \pi\text{E}   \theta _2^{12} \theta
   _3^2 \theta _4^{16}-1120188201990 \pi ^2 \theta _2^{20} \theta
   _4^{12}-19951094609160 \text{E}^2 \theta _2^{16} \theta
   _4^{12}\right.\cr
   &+& \left.9575433427560\pi \text{E}  \theta _2^{16} \theta _3^2
   \theta _4^{12}-61615952970 \pi ^2 \theta _2^{24} \theta
   _4^8-2409808709880 \text{E}^2 \theta _2^{20} \theta
   _4^8\right.\cr
   &+& \left.801748662000 \pi \text{E}  \theta _2^{20} \theta _3^2
   \theta _4^8-6006909480 \text{E}^2 \theta _2^{24} \theta
   _4^4+1520907960\pi \text{E}   \theta _2^{24} \theta _3^2
   \theta _4^4\right.\cr
   &-& \left.1128270 \theta _2^{28} \left(105 \pi ^2 \theta
   _4^4-2764 \text{E}^2\right)\right)\cr
   &+&
   \frac{\pi ^{11} \rho
   ^7}{167382319104000} \left(-2419151723560 \pi
   \theta _4^{28}-8467031032460 \pi  \theta _2^4 \theta
   _4^{24}+4838303447120 \text{E} \theta _3^2 \theta
   _4^{24}\right.\cr
   &-& \left.11188284904940 \pi  \theta _2^8 \theta
   _4^{20}+14514910341360 \text{E} \theta _2^4 \theta _3^2 \theta
   _4^{20}-6803134681200 \pi  \theta _2^{12} \theta
   _4^{16}\right.\cr
   &+& \left.15421216788120 \text{E} \theta _2^8 \theta _3^2 \theta
   _4^{16}-1814356523980 \pi  \theta _2^{16} \theta
   _4^{12}+6650916340640 \text{E} \theta _2^{12} \theta _3^2
   \theta _4^{12}\right.\cr
   &-& \left.151915621000 \pi  \theta _2^{20} \theta
   _4^8+909764632920 \text{E} \theta _2^{16} \theta _3^2 \theta
   _4^8-288182180 \pi  \theta _2^{24} \theta _4^4\right.\cr
   &+& \left.3458186160
   \text{E} \theta _2^{20} \theta _3^2 \theta _4^4-1181803480
   \text{E} \theta _2^{24} \theta _3^2\right) \cr
   &+&
   \frac{\pi ^{11}\rho ^6}{167382319104000} \left(176538062 \theta
   _2^{24}-516584604 \theta _4^4 \theta _2^{20}-135900839598
   \theta _4^8 \theta _2^{16}\right.\cr
   &-& \left.993515335816 \theta _4^{12} \theta
   _2^{12}-2303624732478 \theta _4^{16} \theta _2^8-2168240477484
   \theta _4^{20} \theta _2^4-722746825828 \theta _4^{24}\right)
   \cr
\nonumber
\end{eqnarray}

\section{The coefficients $a_p(\tau)$ in Laurent expansion of Weierstrass function}
\label{Weierstrass}
The Laurent expansion of the Weierstrass function $\wp(z)$ with two periods $\omega_1 =1$ and $\omega_2=\tau$ can be written as
\begin{eqnarray}
\wp(z)&=&\frac{1}{z^2}+\sum_{(n,m)\neq(0,0)}\left[\frac{1}{(z-n-\tau
	m)^2}-\frac{1}{(n+\tau m)^2}\right]\nonumber\\
&=&\frac{1}{z^2}+\sum_{p=2}^{\infty}a_{p}(\tau) z^{2p-2}\nonumber
\end{eqnarray}
where $a_p(\tau)$ are the coefficients in that expansion.
The coefficients $a_p(\tau)$ can all be written
in terms of the elliptic theta functions with the help of the recursion relation (see Ref.\ \cite{Korn} page 749)
\begin{equation}
a_p(\tau)=\frac{3}{(p-3)(2p+1)}~\sum_{k=2}^{p-2}
a_{k}(\tau)a_{p-k}(\tau)\label{aptau}
\end{equation}
where first terms of the sequence are
\begin{eqnarray}
a_2(\tau)&=&\frac{\pi^4}{15}(\theta_2^8+
\theta_2^4\theta_4^4+\theta_4^8)\label{a2tau}\\
a_3(\tau)&=&\frac{\pi^6}{189}(\theta_4^4-\theta_2^4)(\theta_2^4+2\theta_4^4)(\theta_4^4+2\theta_2^4)\label{a3tau}\\
a_4(\tau)&=&\frac{1}{3}a_2^2(\tau)\label{a4tau}\\
a_5(\tau)&=&\frac{3}{11}a_2(\tau)a_3(\tau)\label{a5tau}\\
&\vdots&\nonumber
\end{eqnarray}
Here $\theta_2 \equiv \theta_2(\tau)$,  $\theta_3 \equiv \theta_3(\tau)$,  $\theta_4 \equiv \theta_4(\tau)$ are the elliptic theta functions and in deriving Eqs. (\ref{a2tau}) and (\ref{a3tau}) we have use the following relation
$$
\theta_3^4=\theta_2^4+\theta_4^4
$$
Using the following identities
\begin{center}
	\begin{tabular}{rclcrcl}
		$2\theta_2^2(2\tau)$&$=$&$\theta_3^2-\theta_4^2$&~~~~~~~~~~~~~
		&$\theta_2^2(\tau/2)$&$=$&$2\theta_2\theta_3$\\[0.15cm]
		$2\theta_3^2(2\tau)$&$=$&$\theta_3^2+\theta_4^2$&
		&$\theta_3^2(\tau/2)$&$=$&$\theta_2^2+\theta_3^2$\\[0.15cm]
		$2\theta_4^2(2\tau)$&$=$&$2\theta_3\theta_4$&
		&$\theta_4^2(\tau/2)$&$=$&$\theta_3^2-\theta_2^2$
	\end{tabular}
\end{center}
we can obtained from Eqs. (\ref{a2tau}) and (\ref{a3tau}) that
\begin{eqnarray}
a_2(2\tau)&=&\frac{\pi^4}{15}\frac{\theta_3^8+\theta_4^8
+14\theta_3^4\theta_4^4}{16}\label{a22tau}\\
a_2(\tau/2)&=&\frac{\pi^4}{15}(\theta_3^8+\theta_2^8
+14\theta_3^4\theta_2^4)\label{a2tau12}\\
a_3(2\tau)&=&-\frac{\pi^6}{189}\frac{(\theta_3^4+\theta_4^4)
	(\theta_3^8+\theta_4^8 -34\theta_3^4\theta_4^4)}{32}\label{a32tau}\\
a_3(\tau/2)&=&\frac{\pi^6}{189}2(\theta_3^4+\theta_2^4)
(\theta_3^8+\theta_2^8 -34\theta_3^4\theta_2^4)\label{a3tau12}
\end{eqnarray}
In what follow we will find relations between the coefficients $a_p(2\tau)$, $a_p(\tau)$ and $a_p(\tau/2)$ under transformations $\theta_2 \leftrightarrow \theta_4$ , $\theta_3 \leftrightarrow \theta_4$  and $\theta_2 \leftrightarrow \theta_3$.

Let us start with transformaton $\theta_2 \leftrightarrow \theta_4$. It is easy to see from \eqref{a2tau} - \eqref{a3tau} and Eqs. (\ref{a22tau}) - (\ref{a3tau12}) that under transformation $\theta_2 \leftrightarrow \theta_4$ the coefficients $a_2(2\tau)$, $a_2(\tau)$, $a_2(\tau/2)$, $a_3(2\tau)$, $a_3(\tau)$ and $a_3(\tau/2)$ transforms as
\begin{eqnarray}
a_2(2\tau) &\leftrightarrow& 2^{-4} a_2(\tau/2) \label{a2tau2to12} \\
a_2(\tau) &\leftrightarrow&a_2(\tau) \label{a2tau1to11} \\
a_3(2\tau) &\leftrightarrow& - 2^{-6} a_3(\tau/2) \label{a3tau2to12}\\
a_3(\tau) &\leftrightarrow& -a_3(\tau) \label{a3tau1to1}
\end{eqnarray}
Let us now prove the following Lemma:

Under transformation $\theta_2 \leftrightarrow \theta_4$ the coefficients $a_p(2\tau)$, $a_p(\tau)$ and $a_p(\tau/2)$ transforms as
\begin{eqnarray}
a_p(2\tau) &\leftrightarrow&(-1)^p 2^{-2p} a_p(\tau/2) \label{aptau2to12}\\
a_p(\tau) &\leftrightarrow&(-1)^p a_p(\tau) \label{aptau1to1}
\end{eqnarray}
for any $p$.

Proof: First we can make the following statement:

\noindent (i) for any $p$, right hand side of Eq. (\ref{aptau}) contains terms $a_k a_{p-k}$, where indexes $k$ and $p-k$ are less or equal $p-2$.


Now \eqref{aptau1to1} and \eqref{aptau2to12} can be proved by induction.

Suppose the \eqref{aptau2to12} is true for $p-2$. According to statement (i) the coefficients $a_p(\tau)$ contains only terms with $a_k(\tau)a_{p-k}(\tau)$, where indexes $k$ and $p-k$ are less or equal $p-2$. It is now follow that under transformation $\theta_2 \leftrightarrow \theta_4$ the coefficient $a_p(2\tau)$ transform as
\begin{eqnarray}
a_p(2\tau) &=& \frac{3}{(p-3)(2p+1)}~\sum_{k=2}^{p-2}a_k(2\tau)a_{p-k}(2\tau)
\nonumber\\ &\leftrightarrow& \frac{3}{(p-3)(2p+1)}~\sum_{k=2}^{p-2}(-1)^k 2^{-2k} a_{k}(\tau/2)(-1)^{p-k} 2^{-2(p-k)} a_{p-k}(\tau/2)
\nonumber\\ &=& (-1)^{p} 2^{-2p} a_{p}(\tau/2) \label{aptau2to12v1}
\end{eqnarray}

The \eqref{aptau1to1} can be proved in the similar manner.

Let us now consider behavior of the coefficients $a_p(\tau)$ under transformation $\theta_3 \leftrightarrow \theta_4$. It is easy to see from  \eqref{a2tau} - \eqref{a3tau} and \eqref{a22tau} - \eqref{a3tau12} that under transformation $\theta_3 \leftrightarrow \theta_4$ the coefficients $a_2(2\tau)$, $a_2(\tau)$, $a_2(\tau/2)$, $a_3(2\tau)$, $a_3(\tau)$ and $a_3(\tau/2)$ transforms as
\begin{eqnarray}
a_2(2\tau) &\leftrightarrow&a_2(2\tau)\label{a2tau2to2}\\
a_2(\tau) &\leftrightarrow&a_2(\tau)\label{a2tau1to1}\\
a_2(\tau/2) &\leftrightarrow& (2+2^4) a_2(\tau)-2^4 a_2(2\tau)-a_2(\tau/2) \label{a2tau12toall} \\
a_3(2\tau) &\leftrightarrow&a_3(2\tau)\label{a2tau2to2}\\
a_3(\tau) &\leftrightarrow&a_3(\tau)\label{a2tau2to2}\\
a_3(\tau/2) &\leftrightarrow& (2+2^6) a_3(\tau)-2^6 a_3(2\tau)-a_3(\tau/2) \label{a3tau12toall}
\end{eqnarray}
The general formula for the coefficients $a_p(2\tau)$, $a_p(\tau)$ and $a_p(\tau/2)$ under transformation $\theta_3 \leftrightarrow \theta_4$ are given by
\begin{eqnarray}
a_p(2\tau) &\leftrightarrow&a_p(2\tau)\label{aptau2to2}\\
a_p(\tau) &\leftrightarrow&a_p(\tau)\label{aptau1to1d34}\\
a_p(\tau/2) &\leftrightarrow& (2+2^{2p}) a_p(\tau)-2^{2p} a_p(2\tau)-a_p(\tau/2) \label{aptau12toall}
\end{eqnarray}
The proof is very similar to the proof for above Lemma and we will not give it here.

In the similar way one can proved that behavior of the coefficients  $a_p(2\tau)$, $a_p(\tau)$ and $a_p(\tau/2)$ under transformation $\theta_2 \leftrightarrow \theta_3$ are given by
\begin{eqnarray}
a_p(2\tau) &\leftrightarrow& (1+2^{1-2p}) a_p(\tau)- a_p(2\tau)-2^{-2p}a_p(\tau/2) \label{aptau2toall} \\
a_p(\tau) &\leftrightarrow&a_p(\tau)\label{aptau1to1d23}\\
a_p(\tau/2) &\leftrightarrow&a_p(\tau/2)\label{aptau12to12}
\end{eqnarray}

\section{Relations between $K_{2p}^{1/2, 1/2}(\tau)$, $K_{2p}^{1/2, 0}(\tau)$, $K_{2p}^{0, 1/2}(\tau)$ and $K_{2p}^{0, 0}(\tau)$}
\label{KroneckerToTheta}
In this section we will show that behavior of the ${\rm K}_{2p}^{\frac{1}{2},0}(\tau)$ under transformation $\theta_2 \leftrightarrow \theta_4$ is given by
\begin{equation}
{\rm K}_{2p}^{\frac{1}{2},0}(\tau) \leftrightarrow (-1)^p {\rm K}_{2p}^{0,\frac{1}{2}}(\tau) \label{k2p1001}
\end{equation}
behavior of the ${\rm K}_{2p}^{\frac{1}{2},\frac{1}{2}}(\tau)$ under transformation $\theta_2 \leftrightarrow \theta_3$ is given by
\begin{equation}
{\rm K}_{2p}^{\frac{1}{2},\frac{1}{2}}(\tau) \leftrightarrow {\rm K}_{2p}^{0,\frac{1}{2}}(\tau)\label{k2p1101}
\end{equation}
and behavior of the ${\rm K}_{2p}^{\frac{1}{2},\frac{1}{2}}(\tau)$ under transformation $\theta_3 \leftrightarrow \theta_4$ is given by
\begin{equation}
{\rm K}_{2p}^{\frac{1}{2},\frac{1}{2}}(\tau) \leftrightarrow {\rm K}_{2p}^{\frac{1}{2},0}(\tau)\label{k2p1110}
\end{equation}
The Kronecker functions ${\rm K}_{2p}^{0,0}(\tau)$ can be expressed through  $K_{2p}^{1/2, 1/2}(\tau)$, $K_{2p}^{1/2, 0}(\tau)$, $K_{2p}^{0, 1/2}(\tau)$ as
\begin{eqnarray}
\left(2^{2-2p}-1\right)\,{\rm
	K}^{0,0}_{2p}(\tau)={\rm K}_{2p}^{\frac{1}{2},\frac{1}{2}}(\tau)+{\rm K}_{2p}^{\frac{1}{2},0}(\tau)+{\rm K}_{2p}^{0,\frac{1}{2}}(\tau) \label{K2p111001}
\end{eqnarray}
The Kronecker functions ${\rm K}_{2p}^{0,0}(\tau)$ are related directly to the coefficients $a_p(\tau)$
\begin{equation}
{\rm K}^{0,0}_{2p}(\tau)=
-\frac{(2p)!}{(-4\pi^2)^p}\frac{a_p(\tau)}{(2p-1)}\label{Kabat}
\end{equation}
and as result the Kronecker functions ${\rm K}_{2p}^{0,0}(\tau)$ obeys all equations which satisfied by $a_p(\tau)$, namely \eqref{aptau2to12} - \eqref{aptau1to1} and \eqref{aptau2to2} - \eqref{aptau12to12}.

The Kronecker functions $K_{2p+2}^{\alpha, \beta}(\rho)$ with $(\alpha,\beta)=(1/2,1/2), (1/2,0)$ and $(0,1/2)$ can in turn be related to the function ${\rm K}_{2p}^{0,0}(\tau)$ by means of simple resummation of Kronecker's double series (see for example Appendix F in \cite{Ivasho})
\begin{eqnarray}
{\rm K}_{2p}^{\frac{1}{2},\frac{1}{2}}(\tau)&=& (1+2^{2-2p})\,{\rm
	K}^{0,0}_{2p}(\tau)-2^{1-2p}\,{\rm K}^{0,0}_{2p}(\tau/2)-2\,{\rm
	K}^{0,0}_{2p}(2\tau)\label{K2p11}\\
{\rm K}_{2p}^{\frac{1}{2},0}(\tau)&=& -{\rm
	K}^{0,0}_{2p}(\tau)+2^{1-2p}\,{\rm K}^{0,0}_{2p}(\tau/2)\label{K2p10}\\
{\rm K}_{2p}^{0,\frac{1}{2}}(\tau)&=& - {\rm
	K}^{0,0}_{2p}(\tau)+2\,{\rm
	K}^{0,0}_{2p}(2\tau)\label{K2p01}
\end{eqnarray}
From Eqs. (\ref{K2p11}) - (\ref{K2p01}) one can easily obtain Eq. \eqref{K2p111001}.

Let us now consider behavior ${\rm K}_{2p}^{\frac{1}{2},0}(\tau)$ of under transformation $\theta_2 \leftrightarrow \theta_4$. First we expressed ${\rm K}_{2p}^{\frac{1}{2},0}(\tau)$ in terms of $a_p(\tau)$ and $a_p(\tau/2)$. From \eqref{Kabat} and \eqref{K2p10} we obtain
\begin{eqnarray}
{\rm K}_{2p}^{\frac{1}{2},0}(\tau)&=& -{\rm
	K}^{0,0}_{2p}(\tau)+2^{1-2p}\,{\rm K}^{0,0}_{2p}(\tau/2)\label{K2p10v1}\\
&=&\frac{(2p)!}{(-4\pi^2)^p(2p-1)}\left[a_p(\tau)-2^{1-2p}a_p(\tau/2)\right]
\end{eqnarray}
Now using Eqs. \eqref{aptau2to12} and \eqref{aptau1to1} we can obtain behavior of the ${\rm K}_{2p}^{\frac{1}{2},0}(\tau)$ under transformation $\theta_2 \leftrightarrow \theta_4$
\begin{eqnarray}
{\rm K}_{2p}^{\frac{1}{2},0}(\tau)
&\leftrightarrow&\frac{(2p)!}{(-4\pi^2)^p(2p-1)}\left[(-1)^p\, a_p(\tau)-(-1)^p\, 2\, a_p(\tau/2)\right]\nonumber\\
&=&-(-1)^p\,{\rm K}^{0,0}_{2p}(\tau)+2\,(-1)^p\,{\rm K}^{0,0}_{2p}(\tau/2)\nonumber\\
&=&(-1)^p{\rm K}_{2p}^{0,\frac{1}{2}}(\tau) \label{iden1}
\end{eqnarray}
Thus we have proved Eq. \eqref{k2p1001}. Note, that in derivation of Eq. \eqref{iden1} we have also used Eqs. \eqref{Kabat} and \eqref{K2p01}.

Let us now consider behavior ${\rm K}_{2p}^{\frac{1}{2},\frac{1}{2}}(\tau)$ under transformation $\theta_3 \leftrightarrow \theta_4$. First we expressed ${\rm K}_{2p}^{\frac{1}{2},\frac{1}{2}}(\tau)$ in terms of $a_p(2\tau)$, $a_p(\tau)$ and $a_p(\tau/2)$. From \eqref{Kabat} and \eqref{K2p11} we obtain
\begin{eqnarray}
{\rm K}_{2p}^{\frac{1}{2},\frac{1}{2}}(\tau)&=& (1+2^{2-2p})\,{\rm
	K}^{0,0}_{2p}(\tau)-2^{1-2p}\,{\rm K}^{0,0}_{2p}(\tau/2)-2\,{\rm
	K}^{0,0}_{2p}(2\tau)\label{K2p11v1}\\
&=&-\frac{(2p)!}{(-4\pi^2)^p(2p-1)}\left[(1+2^{2-2p})\,a_p(\tau)-2^{1-2p}\,a_p(\tau/2)-2\,a_p(2\tau)\right]
\end{eqnarray}
Now using Eqs. \eqref{aptau2to2} and \eqref{aptau12toall} we can obtain behavior of the ${\rm K}_{2p}^{\frac{1}{2},\frac{1}{2}}(\tau)$ under transformation $\theta_3 \leftrightarrow \theta_4$
\begin{eqnarray}
{\rm K}_{2p}^{\frac{1}{2},\frac{1}{2}}(\tau)&\leftrightarrow&
\frac{-(2p)!}{(-4\pi^2)^p(2p-1)}\nonumber\\
&\times&\left\{(1+2^{2-2p}) a_p(\tau)
-2^{1-2p}\left[(2+2^{2p}) a_p(\tau)-2^{2p} a_p(2\tau)-a_p(\frac{\tau}{2})\right]-2 a_p(2\tau)\right\}\nonumber\\&=&-\frac{(2p)!}{(-4\pi^2)^p(2p-1)}\left[- a_p(\tau)+2^{1-2p}a_p(\tau/2\right]\nonumber\\
&=&-{\rm K}^{0,0}_{2p}(\tau)+2^{1-2p}\,{\rm K}^{0,0}_{2p}(\tau/2)\nonumber\\
&=&{\rm K}_{2p}^{\frac{1}{2},0}(\tau) \label{iden2}
\end{eqnarray}
Thus we have proved Eq. \eqref{k2p1110}. Along the same lines one can prove the Eq. \eqref{k2p1101}.

\section{Relation between elliptic theta functions, elliptic integral and gamma function}
\label{ThetaToGamma}
The elliptic theta functions and the elliptic integral of the second kind $E$ at particular values
of the aspect ratios $\rho = 1/4, 1/2, 1$ and $2$ given by
\begin{eqnarray}
\theta_2 = 2^{15/16} \sqrt[4]{\frac{2+\sqrt{2}}{\pi }} \sqrt[8]{x}, \hskip 0.4cm
\theta_3 = \left(\sqrt[4]{2}+1\right) \sqrt[4]{\frac{2}{\pi }} \sqrt[8]{x},\hskip 0.4cm
\theta_4 =  \left(\sqrt[4]{2}-1\right) \sqrt[4]{\frac{2}{\pi }} \sqrt[8]{x},\nonumber\\
E = \frac{\sqrt{\pi } \left(\left(11+\sqrt[4]{2}-\sqrt{2}+5\times
   2^{3/4}\right) \sqrt{x}+2
   \left(1+\sqrt[4]{2}\right)^3\right)}{\left(1+\sqrt[4]{2}\right)
   ^3 \left(2+\sqrt{2}+2\times 2^{3/4}\right) \sqrt[4]{x}}
\end{eqnarray}
for $\rho = 1/4$,
\begin{eqnarray}
\theta_2 = \frac{2^{7/8} \sqrt[8]{x}}{\sqrt[4]{\pi }}, \hskip 0.4cm
\theta_3 = \frac{\sqrt{2+\sqrt{2}} \sqrt[8]{x}}{\sqrt[4]{\pi }},\hskip 0.4cm
\theta_4 = \frac{\sqrt{2-\sqrt{2}}
   \sqrt[8]{x}}{\sqrt[4]{\pi }}, \nonumber\\
E = \frac{\sqrt{\pi } \sqrt[4]{x}}{2+\sqrt{2}}+\frac{\sqrt{\pi
   }}{\left(2+\sqrt{2}\right) \sqrt[4]{x}}
\end{eqnarray}
for $\rho = 1/2$,
\begin{eqnarray}
\theta_2 = \theta_4 = \sqrt[4]{\frac{2}{\pi }} \sqrt[8]{x}, \hskip 0.4cm
\theta_3 = \frac{\sqrt{2}}{\sqrt[4]{\pi }} \sqrt[8]{x}, \hskip 0.4cm
E = \frac{1}{4} \sqrt{\pi } \sqrt[4]{\frac{1}{x}}+\frac{\sqrt{\pi }}{2
   \sqrt[4]{\frac{1}{x}}}
\end{eqnarray}
for $\rho = 1$,
\begin{eqnarray}
\theta_2 = \frac{\sqrt{1-\frac{1}{\sqrt{2}}} \sqrt[8]{x}}{\sqrt[4]{\pi }},\hskip 0.4cm
\theta_3 = \frac{\sqrt{1+\frac{1}{\sqrt{2}}} \sqrt[8]{x}}{\sqrt[4]{\pi }},\hskip 0.4cm
\theta_4 =  \frac{2^{3/8} \sqrt[8]{x}}{\sqrt[4]{\pi }},\nonumber\\
E = \frac{\sqrt{\pi } \left(\sqrt{2}
   \sqrt{x}+\frac{1}{2+\sqrt{2}}\right)}{2 \sqrt[4]{x}}
\end{eqnarray}
for $\rho = 2$,
where
\begin{eqnarray}
\nonumber
x = \frac{\pi ^4}{16 \Gamma \left(\frac{3}{4}\right)^8}
\end{eqnarray}
and $\Gamma(z)$ is the gamma function.


\begin{thebibliography}{99}

\bibitem{FisherBarber}
Fisher M and Barber M N 1972 Phys. Rev. Lett. 28 1516
	
\bibitem{Kasteleyn1961}
    P.W. Kasteleyn,  Physica {\bf{27}}, 1209 (1961).


\bibitem{TemFi1961}
    H.N.V Temperley  and M.E. Fisher  ,  Phil. Mag. {\bf{6}}, 1061 (1961).

\bibitem{Brascamp}
H. J. Brascamp and H. Kunz, J. Math. Phys. {\bf 15}, 66 (1974).


\bibitem{Janke}
W. Janke and R. Kenna, Phys. Rev. B {\bf 65}, 064110 (2002).

\bibitem{Izmailian2002} N. Sh. Izmailian, K. B. Oganesyan and C.-K. Hu, Phys. Rev. E {\bf 65}, 056132 (2002).

\bibitem{Ivasho} E. Ivashkevich, N. Sh. Izmailian and Chin-Kun Hu, J. Phys. A: Math. Gen. {\bf 35}, 5543 (2002).

\bibitem{Izmailian2019} N. Sh. Izmailian, Vl. V. Papoyan and R. M. Ziff, J. Phys. A: Math. Theor. {\bf 52}, 335001 (2019).

\bibitem{Weil}
A. Weil, {\it Elliptic functions according to Eisenshtein and
	Kronecker} (Berlin-Heidelberg-New York, Springer-Verlag, 1976).


\bibitem{Izmailian2007} N. Sh. Izmailian and C.-K. Hu, Phys. Rev. E {\bf 76}, 041118 (2007).

\bibitem{asymptotic} V. Kotesovec, On-line encyclopedia of integer sequences, (2013)  (https://oeis.org/A012495).

\bibitem{Korn}
    G. A. Korn and T. M. Korn, {\it Mathematical Handbook} (New-York,
    McGraw-Hill, 1968).



\end{thebibliography}
\end{document}